# Higher dimensional global monopole with cosmological term

F. Rahaman, P.Ghosh, M.Kalam and K. Gayen

Dept. of Mathematics, Jadavpur University,
Kolkata-700032, India

E- Mail: farook_rahaman@yahoo.com

## Abstract:

We investigate the space-time of a global monopole in a five dimensional space-time in presence of the cosmological term. Also the gravitational properties of the monopole solution are discussed.

Pacs Nos. : 98.80cq; 04.20jb; 04.70Bw

## Introduction:

Recent attempts to Unify gravity with other fundamental forces in nature reveal that it is very interesting to study models where the space-time dimension is different from four. Also latest developments in super string and Yang-Mills Super Gravity theory demand, more than usual 4- dimension of space-time. Solutions of Einstein Field equations are believed to of physical relevance possibly at the extremely early times before the Universe under went the compactification transitions [1]. It has also been suggested that the experimental detection of the time variation of fundamental constants could provide strong evidence of the extra dimension [2].

Phase transitions in the early Universe can give rise to topological defects of various kinds such as domain walls, monopoles, strings and their hybrids [3]. In recent, Pando, Valls-Gaboud and Fang [4] have proposed that the topological defects are responsible for structure formation of our Universe. Monopoles are localized defects, will arise if the manifold M contains surfaces, which can't be continuously shrunk to a point i.e. when $\pi_2(M) \neq I$. Global monopoles are important objects both for Particle Physicists and Cosmologists, which predicted to exist in grand unified theory. Using a suitable scalar field it can be shown that spontaneous symmetry breaking can give rise to such objects which are nothing but the topological knots in the vacuum expectation value of the scalar field and most of their energy is concentrated in a small region near the monopole core. In a pioneering work Barriola and Vilenkin (BV) [5] showed that existence of such a monopole solution resulting from the breaking of global S0(3) symmetry of a triplet scalar field in a Schwarzschild background. They found a peculiar result: The space-time produced by a global monopole has no Newtonian gravitational potential in spite of the geometry produced by this heavy object has a non-vanishing curvature.

There has been a fairly large amount of discussions on the gravitational field of global monopole beginning with the work of BV [6]. In recent past, Banerji et al [7] have obtained a monopole solution in Kaluza-Klein space-time, which extends the earlier work of BV to its five dimensional analogue.

The cosmological constant $\Lambda$, which was originally invoked by Einstein to obtain a static solution of his field equations but was subsequently rejected by him after the realization that the Universe is expanding, has fallen in and out of fashion several times. However, recent observations of Type Ia supernovae [8] and measurements of the CMB anisotropy [9] indicating an accelerating Universe have once more drawn attention to the possible existence, at the present epoch, of a positive $\Lambda$-term ( cosmological constant ). It is believed that a positive $\Lambda$-term may dominate the total energy density in the Universe. Observations of high red shift Type Ia supernovae appear to suggest that our Universe may be accelerating with a large fraction of the cosmological density in the form of a cosmological $\Lambda$-term [10]. The confliction of observations with the standard COBE normalized cold dark matter (CMD) model of the structure formation of the Universe with $\Omega_m = 1$ is remedied if the Universe is flat, most of matter smoothly distributed in the form of cosmological constant and only a small fraction $\Omega_m H \approx 0.2$ in the clustered matter ( here H is the Hubble constant in units of 100 km /s/Mpc ). Thus cosmological constant may play a significant role in the structure formation of the Universe, in other words, the $\Lambda$-term might be involved in the formation and evolution of the topological defects. So it is quite justified to study the topological defects with $\Lambda$-term.

Bertand et al[11]has obtained the exact global monopole solutions with cosmological constant in 4d space-time.

As far our knowledge there has not been any work in literature where cosmological term has been considered in higher dimensional space-time of a global monopole.

In this paper, we shall investigate the space-time of the higher dimensional global monopole with cosmological term.

## 2. Field Equations and their integrals:

The metric ansatz describing a monopole in a five dimensional space-time can be written as

$$ds^2 = e^\gamma dt^2 - e^\beta dr^2 - r^2 d\Omega_2^2 - e^\mu d\psi^2 \qquad ....(1)$$

Here $\gamma, \beta, \mu$ are functions of r alone and $\psi$ is the fifth co ordinate.

Here we closely follow the formalism of Banerji et al [7] and take the Lagrangian as

$$L = \tfrac{1}{2} \partial_\mu \Phi^a \partial^\mu \Phi^a - \tfrac{1}{4} \lambda (\Phi^a \Phi^a - \eta^2)^2 \qquad ....(2)$$

where $\Phi^a$ is the triplet scalar field a = 1,2,3 and $\eta$ is the energy scale of symmetry breaking.

The field configuration is taken to be

$$\Phi^a = \eta f(r) (x^a/r) \qquad .....(3)$$

where $x^a x^a = r^2$.

The energy momentum tensors can be written via

$$T_{ab} = 2\,[\partial L / \partial g^{ab}] - L\, g_{ab}$$

$$T_t^{\,t} = \tfrac{1}{2}\,\eta^2\,(f^1)^2\,e^{-\beta} + \eta^2 f^2 r^{-2} + \tfrac{1}{4}\,\lambda\,(\eta^2 f^2 - \eta^2)^2 \qquad \ldots\ldots(4)$$

$$T_r^{\,r} = -\tfrac{1}{2}\,\eta^2\,(f^1)^2\,e^{-\beta} + \eta^2 f^2 r^{-2} + \tfrac{1}{4}\,\lambda\,(\eta^2 f^2 - \eta^2)^2 \qquad \ldots\ldots(5)$$

$$T_\theta^{\,\theta} = T_\varphi^{\,\varphi} = \tfrac{1}{2}\,\eta^2\,(f^1)^2\,e^{-\beta} + \tfrac{1}{4}\,\lambda\,(\eta^2 f^2 - \eta^2)^2 \qquad \ldots\ldots(6)$$

$$T_\psi^{\,\psi} = \tfrac{1}{2}\,\eta^2\,(f^1)^2\,e^{-\beta} + \eta^2 f^2 r^{-2} + \tfrac{1}{4}\,\lambda\,(\eta^2 f^2 - \eta^2)^2 \qquad \ldots\ldots(7)$$

( '1' denotes the differentiation w.r.t. 'r' )

It can be shown that in flat space the monopole core has a size $\delta \sim \sqrt{\lambda}\,\eta^{-1}$ and mass, $M_{core} \sim \lambda^{-1/2}\,\eta$. Thus if $\eta << m_p$ where $m_p$ is the plank mass, it is evident that we can still apply the flat space approximation of $\delta$ and $M_{core}$.
This follows from the fact that in this case the gravity would not much influence on monopole structure.

Banerji et .al assumed that $f = 1$ out side the monopole core [7].
With this result the energy stress tensors assume the following form

$$T_t^{\,t} = T_r^{\,r} = T_\psi^{\,\psi} = (\eta^2 / r^2)\,; \qquad T_\theta^{\,\theta} = T_\varphi^{\,\varphi} = 0 \qquad \ldots\ldots(8)$$

The field equations for the metric (1) reduce to

$$-e^{-\beta}\,[\,\tfrac{1}{2}\,\mu^{11} + \tfrac{1}{4}\,(\mu^1)^2 - \tfrac{1}{4}\,\mu^1 \beta^1 - (\beta^1/r) + (\mu^1/r) + (1/r^2)\,] + (1/r^2)$$
$$= (8\pi G \eta^2 / r^2) + \Lambda \qquad \ldots\ldots(9)$$

$$-e^{-\beta}\,[\,(\mu^1/r) + \tfrac{1}{4}\,\mu^1 \gamma^1 + (\gamma^1/r) + (1/r^2)\,] + (1/r^2)$$
$$= (8\pi G \eta^2 / r^2) + \Lambda \qquad \ldots\ldots(10)$$

$$-e^{-\beta}\,[\,\tfrac{1}{2}\,\gamma^{11} + \tfrac{1}{4}\,(\gamma^1)^2 + (\mu^1/2r) - (\beta^1/2r) + (\gamma^1/2r) + \tfrac{1}{2}\,\mu^{11} + \tfrac{1}{4}\,(\mu^1)^2$$
$$-\tfrac{1}{4}\,\mu^1\beta^1 + \tfrac{1}{4}\,\mu^1\gamma^1 - \tfrac{1}{4}\,\gamma^1\beta^1\,] = \Lambda \qquad \ldots\ldots(11)$$

$$-e^{-\beta}\,[\,\tfrac{1}{2}\,\gamma^{11} + \tfrac{1}{4}\,(\gamma^1)^2 - (\beta^1/r) + (\gamma^1/r) - \tfrac{1}{4}\,\gamma^1\beta^1 + (1/r^2)\,] + (1/r^2)$$
$$= (8\pi G \eta^2 / r^2) + \Lambda \qquad \ldots\ldots(12)$$

For the reason of economy of space we will skip all mathematical details and give the final result as

$$e^{\gamma} = [1 - 8\pi G\eta^2 - (2GM/r) - \tfrac{1}{3} \Lambda r^2]^a$$

$$e^{\beta} = [1 - 8\pi G\eta^2 - (2GM/r) - \tfrac{1}{3} \Lambda r^2]^{-1}$$

$$e^{\mu} = [1 - 8\pi G\eta^2 - (2GM/r) - \tfrac{1}{3} \Lambda r^2]^{(1-a)}$$

…..(13)

Here a is a dimensionless parameter connected with the consistency relation
$a(1-a) = 0$ and M is an integration constant and can be considered as the mass of the monopole core. [ From the (1/r) term in $e^{\gamma}$, it follows that the parameter M is associated with the mass of the configuration i.e. according to BV [5] M having the usual meaning of mass of the central gravitating object ]. It is relevant to point out that unlike BV solution, which is unique, here the criteria of uniqueness, is lost. This result is very similar to Banerji et al [7] solution where the solution is not unique.

## 3. Geodesics:

The five dimensional space-time exterior to the global monopole, is given by (13) reads as

$$ds^2 = A^a dt^2 - A^{-1} dr^2 - r^2 d\Omega_2^2 - A^{(1-a)} d\psi^2 \qquad …(14)$$

where $A = [1 - 8\pi G\eta^2 - (2GM/r) - \tfrac{1}{3} \Lambda r^2]$ …..(15)

We can define a Lagrangian from the metric (14) as

$$L = A^a \dot{t}^2 - A^{-1} \dot{r}^2 - r^2(\dot{\theta}^2 + \sin^2\theta \, \dot{\varphi}^2) - A^{(1-a)} \dot{\psi}^2 \qquad ….(16)$$

where dot represents differentiation with respect to some affine parameter $\xi$. We confine ourselves to motion only in the $\theta = \tfrac{1}{2}\pi$ plane so that the three constants of motion are now

$$g = A^a \dot{t} = \text{constant}$$

$$h = r^2 \dot{\varphi} = \text{constant} \qquad …(17)$$

$$k = A^{(1-a)} \dot{\psi} = \text{constant}$$

where h, k and g are related to angular momentum, five dimensional velocity and energy of the particle respectively.
In general, for photons restricted to the $\theta = \tfrac{1}{2}\pi$ plane equations (16) & (17) lead to the following equation for $a = 1$:

$$(\dot{r}^2 / \dot{\varphi}^2) = (dr/d\varphi)^2 = r^4 [\alpha^2 - (\beta^2 + r^{-2}) A] \qquad …..(18)$$

where $\alpha^2 = (g^2/h^2)$ and $\beta^2 = (k^2/h^2)$.

Following BV's reasoning, we drop the mass term in equation (15), as it is negligible on the astrophysical scale [5].
Hence the equation for the light track is

$$(du/d\varphi)^2 = [B - Du^2 - Cu^4]u^{-2} \qquad \ldots\ldots(19)$$

where $r = (1/u)$ and $D = \beta^2(1 - 8\pi G\eta^2) - \tfrac{1}{3}\Lambda - \alpha^2$;

$B = \tfrac{1}{3}\Lambda\beta^2$; $C = (1 - 8\pi G\eta^2)$

If the fifth dimensional velocity k is large enough so that

$\beta^2 > (\tfrac{1}{3}\Lambda + \alpha^2)(1 - 8\pi G\eta^2)^{-1}$, then the solution of the equation (19) is

$$(1/\sqrt{C})\sin^{-1}[(D + 2Cu^2)/\sqrt{(D^2 + 4BC)}] = 2\varphi$$

For u = 0, one can get

$$2\varphi = (1/\sqrt{C})\sin^{-1}[D/\sqrt{(D^2 + 4BC)}] \qquad \ldots(20)$$

Hence bending of light comes out as

$\pi - 2\varphi$ where $2\varphi$ is given in (20).

Also for the case a = 0, equations (16) & (17) lead to the following equation:

$$(du/d\varphi)^2 = [-C + Bu^2 - Du^4]u^{-2} \qquad \ldots\ldots(21)$$

where $r = (1/u)$ and $B = \alpha^2(1 - 8\pi G\eta^2) - \beta^2$;

$C = \tfrac{1}{3}\Lambda\alpha^2$; $D = (1 - 8\pi G\eta^2)$

If the fifth dimensional velocity k is small enough so that

$\beta^2 < \alpha^2(1 - 8\pi G\eta^2)$, then the solution of the equation (21) is

$$(1/\sqrt{D})\sin^{-1}[(-B + 2Du^2)/\sqrt{(B^2 - 4DC)}] = 2\varphi$$

For u = 0, one can get

$$2\varphi = -(1/\sqrt{D})\sin^{-1}[B/\sqrt{(B^2 - 4DC)}] \qquad \ldots(22)$$

Here $\varphi$ is negative and we refer to it as angular surplus.

## 4. Motion of test particles:

Let us consider a relativistic particle of mass m moving in the gravitational field of the monopole described by eq. (14).
The Hamilton – Jacobi ( H-J) equation is [12]

$$g^{ik} [(\partial S/\delta x^i)][(\partial S/\delta x^k)] + m^2 = 0$$

$$\Rightarrow (1/A)(\partial S/\partial t)^2 - (1/B)(\partial S/\partial r)^2 - (1/r^2)[(\partial S/\partial x_1)^2 + (\partial S/\partial x_2)^2] -$$

$$(1/C)(\partial S/\partial \psi)^2 + m^2 = 0 \qquad \ldots\ldots(23)$$

where

$$A = [1 - 8\pi G\eta^2 - (2GM/r) - \tfrac{1}{3}\Lambda r^2]^a$$

$$B = [1 - 8\pi G\eta^2 - (2GM/r) - \tfrac{1}{3}\Lambda r^2]^{-1}$$

$$C = [1 - 8\pi G\eta^2 - (2GM/r) - \tfrac{1}{3}\Lambda r^2]^{(1-a)}$$

and $x_1, x_2$ are the co ordinates on the surface of the 2 – sphere i.e.
$d\Omega_2^2 \equiv dx_1^2 + dx_2^2 \equiv d\theta^2 + \sin^2\theta \, d\varphi^2$

Take the ansatz $S(t, r, x_1, x_2, \psi) = -E.t + S_1(r) + p_1.x_1 + p_2.x_2 + J.\psi$ …..(24)

as the solution to the H-J eq. (23).

Here the constants E, J are identified as the energy and five dimensional velocity and $p_1, p_2$ are momentum of the particle along different axes on 2 – sphere
with $p = (p_1^2 + p_2^2)^{1/2}$, as the resulting momentum of the particle.
Now substituting (24) in (23), we get

$$S_1(r) = \varepsilon \int [B(E^2/A) - (p^2/r^2) - (J^2/C) + m^2]^{1/2} dr \quad \text{(where } \varepsilon = \pm 1) \quad \ldots..(25)$$

In H-J formalism, the path of the particle is characterized by [12]

$(\partial S/\partial E) = \text{constant}, (\partial S/\partial p_i) = \text{constant} (i = 1,2), (\partial S/\partial J) = \text{constant}$ ……(26)

Thus we get (taking the constants to be zero without any loss of generality),

$$t = \varepsilon \int (\sqrt{BE/A})[(E^2/A) - (p^2/r^2) - (J^2/C) + m^2]^{-1/2} dr \qquad \ldots\ldots(27)$$

$$x_i = \varepsilon \int (\sqrt{B}\, p_i/r^2)[(E^2/A) - (p^2/r^2) - (J^2/C) + m^2]^{-1/2} dr \qquad \ldots..(28)$$

$$\psi = \varepsilon \int (\sqrt{B}J/C)[(E^2/A) - (p^2/r^2) - (J^2/C) + m^2]^{-1/2} dr \qquad \ldots\ldots(29)$$

From (27), we get the radial velocity as

$$(dr/dt) = (A/\sqrt{BE}) [ (E^2/A) - (p^2/r^2) - (J^2/C) + m^2 ]^{1/2} \qquad \ldots\ldots(30)$$

Now the turning points of the trajectory are given by $(dr/dt) = 0$ and as a consequence the potential curves are

$$(E/m) = (A/\sqrt{BE}) [ (E^2/A) - (p^2/r^2) - (J^2/C) + m^2 ]^{1/2} \equiv V(r)$$

In a stationary system E i.e. V(r) must have an extremal value. Hence the value of r for which energy attains it extremal value is given by the equation

$$(dV/dr) = 0 \qquad \ldots\ldots(31)$$

Thus we get the following equations:

**For a = 1**,

$$\tfrac{2}{3}\Lambda (J^2 - m^2) r^5 - 2GM (J^2 - m^2) r^2 + 2p^2 (1 - 8\pi G\eta^2) r - 6GMp^2 = 0 \qquad \ldots(32)$$

This is an algebraic equation of odd degree (degree 5) whose last term is negative. This equation has at least one real positive solution. So it is possible to have bound orbit for the test particle i.e. particle can be trapped by global monopole. In other words, our higher dimensional monopole always exerts gravitational force, which is attractive in nature.

**For a = 0**,

$$[ (4/9) \Lambda^2 p^2 + \tfrac{1}{3} \Lambda J^2 ] r^6 - \tfrac{2}{3}\Lambda(1 - 8\pi G\eta^2) r^4 + [ (4/3)GM\Lambda - 2 GM J^2 ] r^3$$

$$- 2p^2 (1 - 8\pi G\eta^2)^2 r^2 - 4GM(1 - 8\pi G\eta^2) r + 4 (GM)^2 = 0 \qquad \ldots(33)$$

This equation contains so many arbitrary parameters with at least three variations of signs whatever restrictions we imposed (if $(4/3)\Lambda > J^2$, then there are three variations of signs but when $(4/3)\Lambda < J^2$, then there are five variations of signs since $1 - 8\pi G\eta^2 > 0$ as for a typical grand unified theory the parameter $\eta$ is of order $10^{16}$ GeV, so $8\pi G\eta^2 \sim 10^{-5}$). So, by Descartes Rule of Sign, this equation must have at least one positive real roots. So bound orbits are possible i.e. monopole always exerts gravitational force, which is attractive in nature.

## 5. Concluding Remarks:

In this paper we have obtained a class of solutions around a global monopole resulting from breaking of a global S0(3) symmetry in a five dimensional space-time in presence of cosmological term. We see that our higher dimensional monopole metric is not unique like Banerji et al monopole [7]. We have shown that the higher dimensional monopole always exerts gravitational force, which is attractive in nature where as Banerji et al higher dimensional monopole exerts gravitational force provided some restrictions to be imposed [7]. Our higher dimensional monopole, $g_{55}$ loses it dynamical role via a specific choice of an arbitrary constant. We also calculate the bending of light in the above field. For one choice of the parameter, say, $a = 1$, we get a deficit solid angle whereas for other choice, say, $a = 0$, the angular deficit is found to be negative and we refer to it as angular surplus.

## Acknowledgements:

One of the authors, F.R is thankful to IUCAA for providing the research facility.
We are also grateful to the anonymous referee for his/her constructive suggestions.

# References:

[1]   S.Weinberg  Physics in higher dimension  [World Scientific , Singapore, (1986)]

[2]   T. Appelquist, A. Chodas and  P.G.O.  Freund:  Modern Kaluza Klein theories, (Addision Wesley, MA, 1987);  S. Chatterji, Annals of phys. (1992), 218,121   and references there in.

[3]   T.W.B. Kibble, J. Phys. A 9, 1387(1976)
      A.Vilenkin and E.P.S.  Shellard  Cosmic Strings and other Topological Defects (Camb. Univ. Press) (1994)

[4]   J.Pando,D.Valls-Gaboud,and L.Fang Phys.Rev.Lett.81,8568 (1998)

[5]   M.Barriola and A.Vilenkin  Phys. Rev. Lett. 63, 341(1989)

[6]   D. Harrari and C. Lousto Phys. Rev. D 42,2626(1990);
      X. Shi and X. Li Class. Quan. Grav. 8,761(1990);
      P.Breitenlohner, P.Forgacs and D.Maison Nucl.Phys,B383, 357(1992)
      A.Linde astro-ph / 9402031;
      A. Banerji  et al Class. Quan. Grav. 15, 645(1998); I.Cho and A.Vilenkin Phys. Rev. D 56, 7621(1997); S. Chakraborty  Physica Scripta 58,294(1998);
      A. Barros and C. Romero Phys. Rev. D 56,6688 (1997);
      O.Dando and R.Gregory gr-qc/ 9709029; S.Liebling gr-qc/ 9906014;
      gr-qc/ 9904077; E.R.B. de Mello Braz.J.Phys. 31,211(2001);
      R.M.T. Filho and V.B. Bezerra Phys. Rev.D 64,084009 (2001);
      K.A.Bronnikov, B.E.Meierovich and E.R. Podolyak  gr-qc/ 0212091;
      E.R.B. de Mello hep-th/ 0210236;
      X . Li and J. Hao hep-th/ 0210058; K.A.Bronnikov gr-qc/ 0301084

[7]   Banerji.A, S. Chatterji and A. A. Sen Class. Quan. Grav. 13, 3141(1996)

[8]   Perlmutter S et al Astrophys. J 517,565 (1999)

[9]   Caldwell R R  et al Phys.Rev.Lett. 80,1582 (1998); Riess A.G et al Astron. J 116, 1009 (1998)

[10]  V.Sahni and A.Strarobinsky  astro-ph/9904398; Carroll S et al astro-ph/0004075 and  references their in.

[11] Bertand T et al. Class.Quant.Grav.20:4495-4502,(2003)

[12]  S. Chakraborty Gen. Rel. Grav. 28, 1115(1996);
      S. Chakraborty  and L.Biswas  Class. Quan. Grav. 13, 2153 (1996)